# Stochastic Physical Optics & Bell's correlation

*J.F. Geurdes*

C. vd Lijnstraat 164, 2593 NN Den Haag Netherlands

han.geurdes@gmail.com

Abstract:

With the use of classical statistical argumentation similar to the one used in e.g. statistical optics, it is demonstrated that in entanglement of photons, a classical realist explanation cannot be excluded by the CHSH measure in experiment.

## Introduction

Entanglement of photons is a topic of great interest to Quantum Optics. Yet there still remains a nagging possibility that the entanglement is not a quantum effect but can be explained with hidden extra parameters and Stochastic Optical theory in a classical probabilistic framework.

The cornerstone in this debate, as far as experimental results are concerned, comes from the Bell theorem. This theorem purportedly eliminates in experiment all classical Local Realist modeling. In this case, experiments with photonic entanglement appear to be the best testing ground. Time and time again, important research was performed that showed that the CHSH metric can be violated [1].

In a recent paper the author nevertheless demonstrated that the correlation formulated by Bell is possibly unfit to differentiate between quantum and local realist modeling of nature. In the present paper the proof of that point of view is finalized.

## Preliminaries

Starting from the set theoretic result in a previous paper [2] of the author, the question was raised whether or not a local realistic model is actually possible. In the present paper this possibility will be demonstrated using stochastic argumentation. For completeness the author would also like to refer to a draft paper on arXiv. In the latter paper the author showed that from a local hidden variable model the quantum correlation may be obtained. The proposed basic flaw in Bell's reasoning is also indicated in that paper.

The final goal of the present draft is to show that a 2LHV based correlation of Bell can be construed that follows the structure laid down in [2]. The A and B functions under a strict locality condition produce correlations that have a non-zero probability to violate the CHSH. With the stochastics of the example the LHV partitioning of the Omega sets that, according to critics can only be realized with non-local hidden variables, is then established.

## Stochastic reasoning

In the present study the A and B functions are stochastic in certain intervals of the respective hidden variables while in other intervals those functions are not stochastic sign functions. This is sufficient to demonstrate the principle possibility of an LHV explanation within the boundaries of Bell's theorem. It is then demonstrated that Bell's conception of the LHV correlation is unfit to differentiate between quantum and LHV models explaining the violation of the CHSH metric.

In this respect, two aspects of the [2] paper need to be recapitulated. In the first place we have the integral

$$\tfrac{1}{2} P(x,y) = \int_{\lambda \in \Omega_{+|P(a,b)=0}(x,y)} \rho_\lambda - \int_{\lambda \in \Omega_{-|P(a,b)=0}(x,y)} \rho_\lambda \quad (1.1)$$

and the Omega sets defined by

$$\Omega_{0|P(a,b)=0}(x,y) = \{\lambda \in \Lambda \mid A_\lambda(a) B_\lambda(b) = -A_\lambda(x) B_\lambda(y) = \pm 1\} \quad (1.2)$$

and

$$\Omega_{\pm|P(a,b)=0}(x,y) = \{\lambda \in \Lambda \mid A_\lambda(a) B_\lambda(b) = A_\lambda(x) B_\lambda(y) = \pm 1\} \quad (1.3)$$

The plus in the Omega index indicates a +1 for the product while the minus in the Omega index indicates a -1 result for the product.

The previous mentioned integral (1.1) will be treated 'stochastically'. This means we will argue for:

$$\Pr\{P_{LHV}(x,y) = -(x \bullet y)\} > 0 \quad (1.4)$$

Meaning the probability that LHV based Bell correlations exist that violate the CHSH is non-zero.

In the second place the expression of the LHV correlation in (1.1) must match the one presented below for consistency. This is so because we have argued for a partitioning of the $\Lambda$ set into three $\Omega$ sets; $\Omega_{0|P(a,b)=0}(x,y)$, $\Omega_{+|P(a,b)=0}(x,y)$ and $\Omega_{-|P(a,b)=0}(x,y)$. Hence,

$$\tfrac{1}{2} P(x,y) = \int_{\lambda \in \Omega_{0|P(a,b)=0}(x,y)} \rho_\lambda A_\lambda(x) B_\lambda(y) \quad (1.5)$$

## Functions A and B

In this draft the A and B function represent the measurement functions under a strict locality rule. The rule is simply that in no way whatsoever the behavior of the A function can be related to the setting of B and vice versa. In our analysis of [2] we assumed partitioning of the universal set $\Lambda$ of the hidden variables. Hence, we must also reject indirect conveying information through sets that are related to settings. In effect a measurement function may hence only be determined by its settings and its particular hidden variable(s) plus set(s) associated to it.

Let us inspect a two-hidden variable model and associate $\lambda_1$ to A and $\lambda_2$ to B. The universal set $\Lambda$ is a Cartesian product of $\Lambda_1$ for the A side LHV and $\Lambda_2$ for the B side LHV. Hence, $\Lambda = \Lambda_1 \times \Lambda_2$. We subsequently introduce

$$A_{\lambda_1}(x) = \begin{cases} \alpha_{\lambda_1}(x), & for\ \lambda_1 \in I_. \\ \alpha'_{\lambda_1}(x), & for\ \lambda_1 \in \Lambda_1 \setminus I_. \end{cases} \tag{1.6}$$

with, $I_. \subset \Lambda_1$ partitioning $\Lambda_1$ into two sets. Similarly for the B function it obtains:

$$B_{\lambda_2}(y) = \begin{cases} \beta_{\lambda_2}(y), & for\ \lambda_2 \in J_. \\ \beta'_{\lambda_2}(y), & for\ \lambda_2 \in \Lambda_2 \setminus J_. \end{cases} \tag{1.7}$$

Moreover, $J_. \subset \Lambda_2$ creates a partitioning of $\Lambda_2$ into two sets. The $\alpha$ and $\beta$ functions project in the set $\{-1, 1\}$. In addition, the dot denotes the place where an index value can be filled in (see the selection rule below).

It is supposed that proper $A_{\lambda_1}(a)$ and $B_{\lambda_2}(b)$ always can be obtained for convenient orthogonal A and B settings such that Omega sets can be construed with the definitions in (1.6) and (1.7). Later on we will have more to say about this matter.

## Probability density of the hidden variables

In our example let us postulate a density for $(\lambda_1, \lambda_2) \in \left[-\frac{1}{\sqrt{2}}, \frac{1}{\sqrt{2}}\right] \times \left[-\frac{1}{\sqrt{2}}, \frac{1}{\sqrt{2}}\right] \equiv \Lambda$. Here, $[p, q] = \{x \in \mathbb{R} \mid p \leq x \leq q\}$. The density is (for $n = 1, 2$)

$$\rho_{\lambda_n} = \begin{cases} \frac{1}{\sqrt{2}}, & \lambda_n \in \left[-\frac{1}{\sqrt{2}}, \frac{1}{\sqrt{2}}\right] \\ 0, & elsewhere \end{cases} \tag{1.8}$$

## Form of the integrals in the example 2LHV density

Using the density we now may write for this two LHV model

$$P(x, y) = \iint_{(\lambda_1,\lambda_2) \in \Omega_{+|P(a,b)=0}(x,y)} d\lambda_1 d\lambda_2 - \iint_{(\lambda_1,\lambda_2) \in \Omega_{-|P(a,b)=0}(x,y)} d\lambda_1 d\lambda_2 \qquad (1.9)$$

Moreover, the consistency condition in (1.5) now becomes

$$P(x, y) = \iint_{(\lambda_1,\lambda_2) \in \Omega_{0|P(a,b)=0}(x,y)} A_{\lambda_1}(x) B_{\lambda_2}(y) d\lambda_1 d\lambda_2 \qquad (1.10)$$

# Parameter selection rules

In the example for the ease of presentation the following is supposed. Let us select, $1_A = (1,0,0)$ and $2_A = (0,1,0)$ for the A wing of the experiment. For the B wing let us take $1_B = \left(\frac{1}{\sqrt{2}}, \frac{-1}{\sqrt{2}}, 0\right)$ and $2_B = \left(\frac{-1}{\sqrt{2}}, \frac{-1}{\sqrt{2}}, 0\right)$. The indices on the numbers indicate the wing. The number notation is a short hand for indicating the parameter vectors. Note e.g. $a = (0, \frac{-1}{\sqrt{2}}, \frac{1}{\sqrt{2}})$ and $b = (0, \frac{1}{\sqrt{2}}, \frac{1}{\sqrt{2}})$ and $a \cdot b = 0$ to match $P(a,b) = 0$ and $x$ and $y$ unequal to $a$ and $b$.

In [2] it was described how settings of a measurement instrument restrict the $\lambda_1$ and $\lambda_2$ intervals. In the two tables below this selection rule is recapitulated. For the A instrument:

| Setting | Interval |
|---|---|
| $1_A$ | $I_1 = \left[\frac{-1}{\sqrt{2}}, 1 - \frac{1}{\sqrt{2}}\right] \subset \left[\frac{-1}{\sqrt{2}}, \frac{1}{\sqrt{2}}\right]$ |
| $2_A$ | $I_2 = \left[-1 + \frac{1}{\sqrt{2}}, \frac{1}{\sqrt{2}}\right] \subset \left[\frac{-1}{\sqrt{2}}, \frac{1}{\sqrt{2}}\right]$ |

For the B instrument:

| Setting | Interval |
|---|---|
| $1_B$ | $J_1 = \left[\frac{-1}{\sqrt{2}}, 0\right] \subset \left[\frac{-1}{\sqrt{2}}, \frac{1}{\sqrt{2}}\right]$ |
| $2_B$ | $J_2 = \left[0, \frac{1}{\sqrt{2}}\right] \subset \left[\frac{-1}{\sqrt{2}}, \frac{1}{\sqrt{2}}\right]$ |

In terms of measurement one can imagine a first 'stage gate' where the selection rule in the two tables is implemented. In a subsequent 'stage gate' the proper A and B response function, given in (1.6) and (1.7), are implemented. If the entrance of the to-be-measured particle is the second 'stage gate' then the selection of the A and B function in the measurement instrument is also affected by the event of the entrance of the particle. Hence, third stage gate A and B function selection will warrant proper local stochastic behavior for the functions.

In order to violate the CHSH, four pairs, $(1_A,1_B)$, $(1_A,2_B)$, $(2_A,1_B)$ and $(2_A,2_B)$ need to be inspected on their LHV correlation and consistency with the idea of partitioning. If

$$\Pr\{|P_{LHV}(1_A,1_B) - P_{LHV}(1_A,2_B) - P_{LHV}(2_A,1_B) - P_{LHV}(2_A,2_B)| > 2\} > 0 \qquad (1.11)$$

then Bell's expression for correlation is unfit for discriminating between all LHV models and quantum mechanical result.

## Setting pair $(1_A,1_B)$

### 1. The correlation

As can be seen from the definitions of $1_A$ and $1_B$ the quantum correlation in this case is equal to $P_{QM}(1_A,1_B) = -\dfrac{1}{\sqrt{2}}$. The LHV expression is expected to behave like presented in (1.4).

From the selection rule in the previous two tables we may obtain, in the first stage gate, the partitioning into the set $I_1 \times J_1$ and the set $\Lambda \setminus (I_1 \times J_1)$. Note that $\Lambda = \left[\dfrac{-1}{\sqrt{2}}, \dfrac{1}{\sqrt{2}}\right] \times \left[\dfrac{-1}{\sqrt{2}}, \dfrac{1}{\sqrt{2}}\right]$.

Let us suppose that the $\alpha_{\lambda_1}(x)$ from (1.6) and the $\beta_{\lambda_2}(y)$ from (1.7) are $\pm 1$ signs obtained from tossing fair (but separated) coins. Let us denote this with $\alpha_{\lambda_1}(x) = F_{ct}(1)$ and $\beta_{\lambda_2}(y) = F_{ct}(2)$. Under the convenient assumption that $A_{\lambda_1}(a)$ and $B_{\lambda_2}(b)$ always can be obtained for suitable orthogonal A and B settings we find in probabilistic terms that

$$\Pr\{\Omega_{+|P(a,b)=0}(1_A,1_B) = \varnothing \ \& \ \Omega_{-|P(a,b)=0}(1_A,1_B) = I_1 \times J_1\} > 0 \qquad (1.12)$$

This is so because with $\alpha_{\lambda_1}(x) = F_{ct}(1)$ and $\beta_{\lambda_2}(y) = F_{ct}(2)$ there is a non-zero probability that, under convenient $A_{\lambda_1}(a)$ and $B_{\lambda_2}(b)$, $A_{\lambda_1}(1_A)B_{\lambda_2}(1_B) = -1$.

Note that if we drop the convenient $A_{\lambda_1}(a)$ and $B_{\lambda_2}(b)$ assumption and make $A_{\lambda_1}(a)$ and $B_{\lambda_2}(b)$ stochastic too the probability in (1.12) will be smaller but still nonzero. The latter fact is what counts so it is allowed to argue under convenient $A_{\lambda_1}(a)$ and $B_{\lambda_2}(b)$. Note that, $P(a,b) = 0$, the $\Omega_0$ set implies $A_{\lambda_1}(a)B_{\lambda_2}(b) = -A_{\lambda_1}(x)B_{\lambda_2}(y)$ and in the $\Omega_\pm$ sets $A_{\lambda_1}(a)B_{\lambda_2}(b) = A_{\lambda_1}(x)B_{\lambda_2}(y)$.

If it is subsequently observed that (1.12) implies that

$$\Pr\left\{\iint_{(\lambda_1,\lambda_2)\in\Omega_{+|P(a,b)=0}(1_A,1_B)} d\lambda_1 d\lambda_2 - \iint_{(\lambda_1,\lambda_2)\in\Omega_{-|P(a,b)=0}(1_A,1_B)} d\lambda_1 d\lambda_2 = -\dfrac{1}{\sqrt{2}}\right\} > 0 \qquad (1.13)$$

then we see $\Pr\left\{P(1_A,1_B)=-\dfrac{1}{\sqrt{2}}\right\}>0$.

## 2. The consistency condition

For the sake of easy argumentation let us introduce the notation $P(1_A,1_B)|_{\Omega_\pm}$ for the LHV correlation based on the form in (1.9). Moreover, $P(1_A,1_B)|_{\Omega_0}$ is the LHV correlation based on the form (1.10). In our case the consistency condition is: $P(1_A,1_B)|_{\Omega_0}$ is equal to $P(1_A,1_B)|_{\Omega_\pm}$.

When either the stochastic assignment leads to $\Omega_{+|P(a,b)=0}(1_A,1_B)=\varnothing$ & $\Omega_{-|P(a,b)=0}(1_A,1_B)=I_1\times J_1$ or $\Omega_{-|P(a,b)=0}(1_A,1_B)=\varnothing$ & $\Omega_{+|P(a,b)=0}(1_A,1_B)=I_1\times J_1$ we see that, under convenient $A_{\lambda_1}(a)$ and $B_{\lambda_2}(b)$, $\Omega_0$ is equal to

$$\Omega_{0|P(a,b)=0}(1_A,1_B)=\Lambda\setminus(I_1\times J_1) \tag{1.14}$$

Reformulating $\Omega_0$ into a more explicit form this leads to

$$\Omega_{0|P(a,b)=0}(1_A,1_B)=(I_1\times J_2)\cup((\Lambda_1\setminus I_1)\times J_1)\cup((\Lambda_1\setminus I_1)\times J_2) \tag{1.15}$$

Recall that $\Lambda\equiv\Lambda_1\times\Lambda_2$ with $\Lambda_1=\Lambda_2=\left[\dfrac{-1}{\sqrt{2}},\dfrac{1}{\sqrt{2}}\right]$ and the sets $I_1, J_1$ and $J_2$ as defined in the selection rule tables.

From the definition of the A and B functions in (1.6) and (1.7) and $\alpha_{\lambda_1}(x)=F_{ct}(1)=\alpha$ together with $\beta_{\lambda_2}(y)=F_{ct}(2)=\beta$ brings us to the following expression for $P(1_A,1_B)|_{\Omega_0}$

$$P(1_A,1_B)|_{\Omega_0}=\iint\limits_{(\lambda_1,\lambda_2)\in I_1\times J_2}\alpha\beta'_{\lambda_2}(1_B)d\lambda_1 d\lambda_2+\iint\limits_{(\lambda_1,\lambda_2)\in(\Lambda_1\setminus I_1)\times J_1}\alpha'_{\lambda_1}(1_A)\beta d\lambda_1 d\lambda_2+\iint\limits_{(\lambda_1,\lambda_2)\in(\Lambda_1\setminus I_1)\times J_2}\alpha'_{\lambda_1}(1_A)\beta'_{\lambda_2}(1_B)d\lambda_1 d\lambda_2 \tag{1.16}$$

This is true because for e.g. $\lambda_1\in I_1$ we have $A_{\lambda_1}(1_A)=\alpha_{\lambda_1}(1_A)=\alpha\in\{-1,1\}$ based on a fair coin toss $F_{ct}$. Similarly for $\lambda_2\in J_2$ we have $B_{\lambda_2}(1_B)=\beta'_{\lambda_2}(1_B)\in\{-1,1\}$ while for $\lambda_2\in J_1$ we see that $B_{\lambda_2}(1_B)=\beta_{\lambda_2}(1_B)=\beta\in\{-1,1\}$ etc.

Because of the consistency condition we need to show that: $P(1_A,1_B)|_{\Omega_0} = P(1_A,1_B)|_{\Omega_\pm} = \frac{-1}{\sqrt{2}}$. Let us introduce the following notation. We have

$$U = \int_{\lambda_2 \in J_2} \beta'_{\lambda_2}(1_B) d\lambda_2 \tag{1.17}$$

together with

$$V = \int_{\lambda_1 \in \Lambda_1 \setminus I_1} \alpha'_{\lambda_1}(1_A) d\lambda_1 \tag{1.18}$$

Because, $\int_{\lambda_1 \in I_1} d\lambda_1 = \left(1 - \frac{1}{\sqrt{2}}\right) - \left(\frac{-1}{\sqrt{2}}\right) = 1$ and $\int_{\lambda_2 \in J_1} d\lambda_2 = 0 - \left(\frac{-1}{\sqrt{2}}\right) = \frac{1}{\sqrt{2}}$ the following algebraic equation occurs after multiplying (1.16) on both sides with $\alpha$

$$U - \tfrac{1}{\sqrt{2}} V + \alpha UV = \tfrac{-\alpha}{\sqrt{2}} \tag{1.19}$$

Here, $\alpha\beta = -1$, is employed because of the necessity to have a nonzero probability for $\Omega_{+|P(a,b)=0}(1_A,1_B) = \emptyset$ & $\Omega_{-|P(a,b)=0}(1_A,1_B) = I_1 \times J_1$. The definitions in (1.17) and (1.18) are used that lead us to (1.19).

3. *Numerics of the consistency condition on partitioning*

The equation in (1.19) appears to be difficult to solve exactly. In this section the existence of $\alpha'_{\lambda_1}(1_A)$ and $\beta'_{\lambda_2}(1_B)$ will be shown by approximately solve the U,V equation. To this end let us select mathematical forms for $\alpha'_{\lambda_1}(1_A)$ and $\beta'_{\lambda_2}(1_B)$.

Suppose, for proper, $\eta(1_B)$ and $\zeta(1_A)$ that

$$\exists_{\eta(1_B) \in J_2} \beta'_{\lambda_2}(1_B) = sign\left(\eta(1_B) - \lambda_2\right) \tag{1.20}$$

For completeness: $J_2 = \left[0, \tfrac{1}{\sqrt{2}}\right]$. As can be verified we may write

$$U = \int_0^{\eta(1_B)} d\lambda_2 - \int_{\eta(1_B)}^{\frac{1}{\sqrt{2}}} d\lambda_2 = 2\eta(1_B) - \frac{1}{\sqrt{2}} \tag{1.21}$$

Hence, $U \in \left[\tfrac{-1}{\sqrt{2}}, \tfrac{1}{\sqrt{2}}\right]$. In addition, suppose

$$\exists_{\zeta(1_A) \in \Lambda_1 \setminus I_1} \alpha'_{\lambda_1}(1_A) = sign(\zeta(1_A) - \lambda_1) \tag{1.22}$$

For completeness: $\Lambda_1 \setminus I_1 = \left[1-\frac{1}{\sqrt{2}}, \frac{1}{\sqrt{2}}\right]$. As can be verified we also have

$$V = \int_{1-\frac{1}{\sqrt{2}}}^{\zeta(1_A)} d\lambda_1 - \int_{\zeta(1_A)}^{\frac{1}{\sqrt{2}}} d\lambda_1 = 2\zeta(1_A) - 1 + \frac{1}{\sqrt{2}} - \frac{1}{\sqrt{2}} = 2\zeta(1_A) - 1 \qquad (1.23)$$

This implies that in numerical exploration of possible zeroes of (1.19) we see $V \in \left[1-\sqrt{2}, \sqrt{2}-1\right]$.

With the interval conditions $U \in \left[\frac{-1}{\sqrt{2}}, \frac{1}{\sqrt{2}}\right]$ and $V \in \left[1-\sqrt{2}, \sqrt{2}-1\right]$ and exploring $\alpha = 1$ the following numerical results were found for $\delta(U,V) = \left|U - \frac{1}{\sqrt{2}}V + \alpha UV - \left(\frac{-\alpha}{\sqrt{2}}\right)\right|$

| U | V | Step size h | $\delta(U,V)$ |
|---|---|---|---|
| -0.60711 | 0.075786 | 0.01 | 0.0004 |
| -0.45511 | 0.215786 | 0.001 | 9.1x10$^{-6}$ |
| -0.45371 | 0.218186 | 0.0001 | 9.9x10$^{-7}$ |

Similarly for $\alpha = -1$ we found

| U | V | Step size h | $\delta(U,V)$ |
|---|---|---|---|
| 0.31000 | -0.40421 | 0.01 | 2.1x10$^{-5}$ |
| 0.32300 | -0.37421 | 0.001 | 4.7x10$^{-6}$ |
| 0.32760 | -0.36691 | 0.0001 | 8.0x10$^{-7}$ |

The numerical exactness of the approximation can of course always be better. For the first row in the $\alpha = -1$ case the following plot of the error $\delta(U,V)$ can be shown.

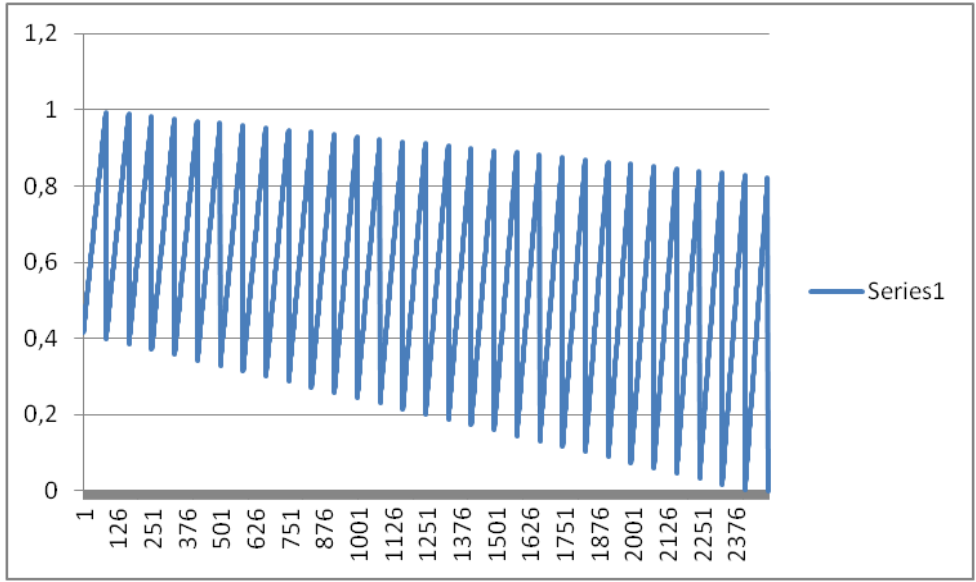

**Figure 1** Plot of error term in iteration step (step size 0.01 and max error 1x10$^{-4}$).

Hence, there exists a solution to (1.19) 'in principle' within the boundaries of the approximate open intervals $U \in (-0{,}70711, 0{,}70711)$ and $V \in (-0{,}414214, 0{,}414214)$. This implies that both $\zeta(1_A)$ and $\eta(1_B)$ are possible 'in principle'. Hence, $\alpha'_{\lambda_1}(1_A)$ and $\beta'_{\lambda_2}(1_B)$ can be obtained when $\Omega_{+|P(a,b)=0}(1_A,1_B) = \varnothing$ & $\Omega_{-|P(a,b)=0}(1_A,1_B) = I_1 \times J_1$.

## 4. Concluding remarks on correlation and consistency for $(1_A, 1_B)$

In the previous paragraphs of this section it was demonstrated that $\Pr\left\{P(1_A,1_B)|_{\Omega_\pm} = -\frac{1}{\sqrt{2}}\right\} > 0$ in a manner that is consistent with partitioning of the $\Lambda$ set of LHVs, or, because of numerical demonstration only: $\Pr\left\{P(1_A,1_B)|_{\Omega_0} \approx -\frac{1}{\sqrt{2}}\right\} > 0$. This allows the conclusion that (misusing slightly the Bayesian condition notation)

$$\Pr\left\{P_{LHV}(1_A,1_B) = -\tfrac{1}{\sqrt{2}} \big| Consistency\right\} > 0 \qquad (1.24)$$

The consistency refers to the partitioning of the $\Lambda$ set of LHVs without the need to renormalize the probability density function defined in equation (1.8).

# Setting pair $(1_A, 2_B)$

## 1. Correlation and stochastics

In the case of setting pair, $(1_A, 2_B)$ the quantum correlation is $P_{QM}(1_A, 2_B) = \frac{1}{\sqrt{2}}$. From the selection rule in the previous sections the parameter settings and LHV intervals are determined. For $\lambda_1$ we have $I_1 = \left[\frac{-1}{\sqrt{2}}, 1 - \frac{1}{\sqrt{2}}\right]$ while for $\lambda_2$ the set $J_2 = \left[0, \frac{1}{\sqrt{2}}\right]$ is selected in the first stage gate of the instrument.

Let us subsequently use the function definitions in (1.6) and (1.7) and have in the third stage gate (i.e. after the particle hit the measurement area of the measuring instrument) $\alpha_{\lambda_1}(x) = F_{ct}(1) = \alpha$ together with $\beta_{\lambda_2}(y) = F_{ct}(2) = \beta$. The stochastics may lead us to the following conclusion, under convenient $A_{\lambda_1}(a)$ and $B_{\lambda_2}(b)$

$$\Pr\{\Omega_{+|P(a,b)=0}(1_A, 2_B) = I_1 \times J_2 \ \& \ \Omega_{-|P(a,b)=0}(1_A, 2_B) = \varnothing\} > 0 \tag{1.25}$$

Hence, $\Pr\left\{P(1_A, 2_B)|_{\Omega_{\pm}} = \frac{1}{\sqrt{2}}\right\} > 0$ in a similar manner as was detailed in the previous section.

## 2. Consistency and numerics

The consistency can be derived from the $\Omega_0$ set

$$\Omega_{0|P(a,b)=0}(1_A, 2_B) = (I_1 \times J_1) \cup ((\Lambda_1 \setminus I_1) \times J_2) \cup ((\Lambda_1 \setminus I_1) \times J_1) \tag{1.26}$$

The associated expression for $P(1_A, 2_B)|_{\Omega_0}$ then follows

$$P(1_A, 2_B)|_{\Omega_0} = \iint\limits_{(\lambda_1, \lambda_2) \in I_1 \times J_1} \alpha \beta'_{\lambda_2}(2_B) d\lambda_1 d\lambda_2 + \iint\limits_{(\lambda_1, \lambda_2) \in (\Lambda_1 \setminus I_1) \times J_2} \alpha'_{\lambda_1}(1_A) \beta d\lambda_1 d\lambda_2 + \iint\limits_{(\lambda_1, \lambda_2) \in (\Lambda_1 \setminus I_1) \times J_1} \alpha'_{\lambda_1}(1_A) \beta'_{\lambda_2}(2_B) d\lambda_1 d\lambda_2 \tag{1.27}$$

Let us subsequently define the $U$ and $V$ for $(1_A, 2_B)$ using a similar form for $\alpha'_{\lambda_1}(1_A)$ and $\beta'_{\lambda_2}(1_B)$ as was used in the previous section, with

$$U = \int\limits_{\lambda_2 \in J_1} \beta'_{\lambda_2}(2_B) d\lambda_2 \tag{1.28}$$

and

$$V = \int\limits_{\lambda_1 \in \Lambda_1 \setminus I_1} \alpha'_{\lambda_1}(1_A) d\lambda_1 \tag{1.29}$$

As was argued previously we have $U \in \left[\frac{-1}{\sqrt{2}}, \frac{1}{\sqrt{2}}\right]$ and $V \in \left[1-\sqrt{2}, \sqrt{2}-1\right]$ as can be easily checked.

Because $\alpha\beta = 1$ is under study and $P_{QM}(1_A, 2_B) = \frac{1}{\sqrt{2}}$, the $U, V$ equation becomes

$$U + \tfrac{1}{\sqrt{2}}V + \alpha UV = \tfrac{\alpha}{\sqrt{2}} \qquad (1.30)$$

Using the numerical precision defined by $\delta(U,V) = \left|U + \tfrac{1}{\sqrt{2}}V + \alpha UV - \left(\tfrac{\alpha}{\sqrt{2}}\right)\right|$ we found for $\alpha = -1$

| U | V | Step size h | $\delta(U,V)$ |
|---|---|---|---|
| -0.67711 | -0.0142 | 0.01 | $1.8 \times 10^{-4}$ |
| -0.67711 | -0.0217 | 0.0001 | $4.1 \times 10^{-5}$ |
| -0.67710 | -0.0216 | 0.00001 | $8.0 \times 10^{-7}$ |

For $\alpha = 1$ we found

| U | V | Step size h | $\delta(U,V)$ |
|---|---|---|---|
| 0.3700 | 0.3258 | 0.01 | $5.4 \times 10^{-4}$ |
| 0.3001 | 0.4042 | 0.0001 | $3.4 \times 10^{-5}$ |
|  |  |  |  |

Hence, there exists a solution to (1.30) 'in principle' within the boundaries of the approximate open intervals $U \in (-0{,}70711, 0{,}70711)$ and $V \in (-0{,}414214, 0{,}414214)$.

### 3. Concluding remarks on correlation and consistency for $(1_A, 2_B)$

In a numerical approximation it was found that the consistency: $\Pr\left\{P(1_A, 2_B)|_{\Omega_0} \approx \frac{1}{\sqrt{2}}\right\} > 0$ can be obtained. Hence, for $(1_A, 2_B)$ we also may conclude that

$$\Pr\left\{P(1_A, 2_B) = \tfrac{1}{\sqrt{2}} \big| Consistency\right\} > 0 \qquad (1.31)$$

## Conclusion

From the numerical analysis on consistency we may draw the conclusion that not only for the pair $(1_A, 1_B)$, $\Pr\left\{P_{LHV}(1_A, 1_B) = -\tfrac{1}{\sqrt{2}} \big| Consistency\right\} > 0$, but also for $(1_A, 2_B), (2_A, 2_B)$ and $(2_A, 1_B)$ pairs we may see,

$$\Pr\{P_{LHV}(1_A, 2_B) = \tfrac{1}{\sqrt{2}} | Consistency\} > 0,$$
$$\Pr\{P_{LHV}(2_A, 2_B) = \tfrac{1}{\sqrt{2}} | Consistency\} > 0, \quad (1.32)$$
$$\Pr\{P_{LHV}(2_A, 1_B) = \tfrac{1}{\sqrt{2}} | Consistency\} > 0$$

Please do observe that the 'stage gating' of the measurement instrument such as described in the previous sections implies that the event of a particle hitting the 'measurement area' of the instrument ensures proper randomization of function response. Meaning that in reference to (1.32) $\alpha_{\lambda_1}(x)$ and $\beta_{\lambda_2}(y)$ in their proper I and J sets are stochastic random projections in $\{-1,1\}$, with $\alpha\beta = 1$ such that $\Pr\{\Omega_{-|P(a,b)=0}(x,y) = \varnothing \ \& \ \Omega_{+|P(a,b)=0}(x,y) = I(x) \times J(y)\} > 0$ for $(x,y)$ equal to $(1_A, 2_B), (2_A, 2_B)$ or $(2_A, 1_B)$. The functions $\alpha'_{\lambda_1}(1_A)$ and $\beta'_{\lambda_2}(1_B)$ are adapted to them. The respective $I(x) \times J(y)$ are determined according to the selection rule in the selection rule tables.

In (1.25) the consistency conditions reflect the partitioning of the set $\Lambda$ without the necessity to renormalize the probability density in (1.8). This finally lead us to the already anticipated conclusion in [2] that there is a non-zero probability, small as it may be, such that LHVs violate the CHSH metric. We would also like to stress that strict locality is obeyed in all steps of the computations and in all stage gates of the separate measurement instruments.

Hence, the conclusion of the author in [2] that Bell's correlation is an unfit measure to differentiate between LHV and quantum non-locality in experiment is supported with a stochastic argument.

This conclusion is related to the one expressed by Joy Christian in an arXiv publication [3]. The present author believes that this paper of Christian firmly holds true because previous attacks on it were rightfully rejected. Note that the claim of the author also is supported by earlier publications such as [4].